\documentclass{svproc}

\usepackage{amsmath,amssymb,amsfonts}
\usepackage{algorithmic}
\usepackage{graphicx}
\usepackage{textcomp}
\usepackage{xcolor}
\usepackage{fancyhdr}

\def\BibTeX{{\rm B\kern-.05em{\sc i\kern-.025em b}\kern-.08em
    T\kern-.1667em\lower.7ex\hbox{E}\kern-.125emX}}
\usepackage[nospace]{cite}
\graphicspath{ {./images/} }
\usepackage{url}
\usepackage{color}
\usepackage{float}

\begin{document}

\title{A Survey of 6G Wireless Communications: Emerging Technologies}
\titlerunning{A Survey of 6G Wireless Communications: Emerging Technologies}

\author{Yang Zhao\inst{1} \and  Jun Zhao\inst{1} \and Wenchao Zhai\inst{2} \and Sumei Sun\inst{3,4} \and Dusit Niyato\inst{1} \and Kwok-Yan Lam\inst{1}}
\authorrunning{ } 
%
%
\institute{School of Computer Science and Engineering, Nanyang Technological University, Singapore,\\
\and
College of Information Engineering, Jiliang University, Hangzhou, China,
\and
Institute for Infocomm Research (I2R), Agency for Science, Technology and Research (A*STAR), Singapore,\\ 
\and
Infocomm Technology Cluster, Singapore Institute of Technology, Singapore,
\email{ $^1$\{s180049@e., junzhao, dniyato, kwokyan.lam\}@ntu.edu.sg, $^2$zhaiwenchao@cjlu.edu.cn, $^{3,4}$sunsm@i2r.a-star.edu.sg}
}

\maketitle
\thispagestyle{fancy}
\pagestyle{fancy}
\lhead{This paper appears in the Proceedings of Future of Information and Communication Conference (FICC) 2021.}
\cfoot{\thepage}
\renewcommand{\headrulewidth}{0.4pt}
\renewcommand{\footrulewidth}{0pt}

\begin{abstract}
While fifth-generation (5G) communications are being rolled out around the world, sixth-generation (6G) communications have attracted much attention from both the industry and the academia. Compared with 5G, 6G will have a wider frequency band, higher transmission rate, spectrum efficiency, greater connection capacity, shorter delay, wider coverage and stronger anti-interference capability, so as to meet the various network requirements for industries. In this paper, we present a survey of potential essential technologies in 6G. In particular, we will introduce index modulation, artificial intelligence, intelligent surfaces, and terahertz communications technologies in detail, while giving a brief introduction to other potential technologies, including visible light communications, blockchain-enabled wireless network, advanced duplex, holographic radio and network in box. 
\end{abstract}


\section{Introduction}
\begin{figure}[!h]
    \centering
     \includegraphics[scale=0.56]{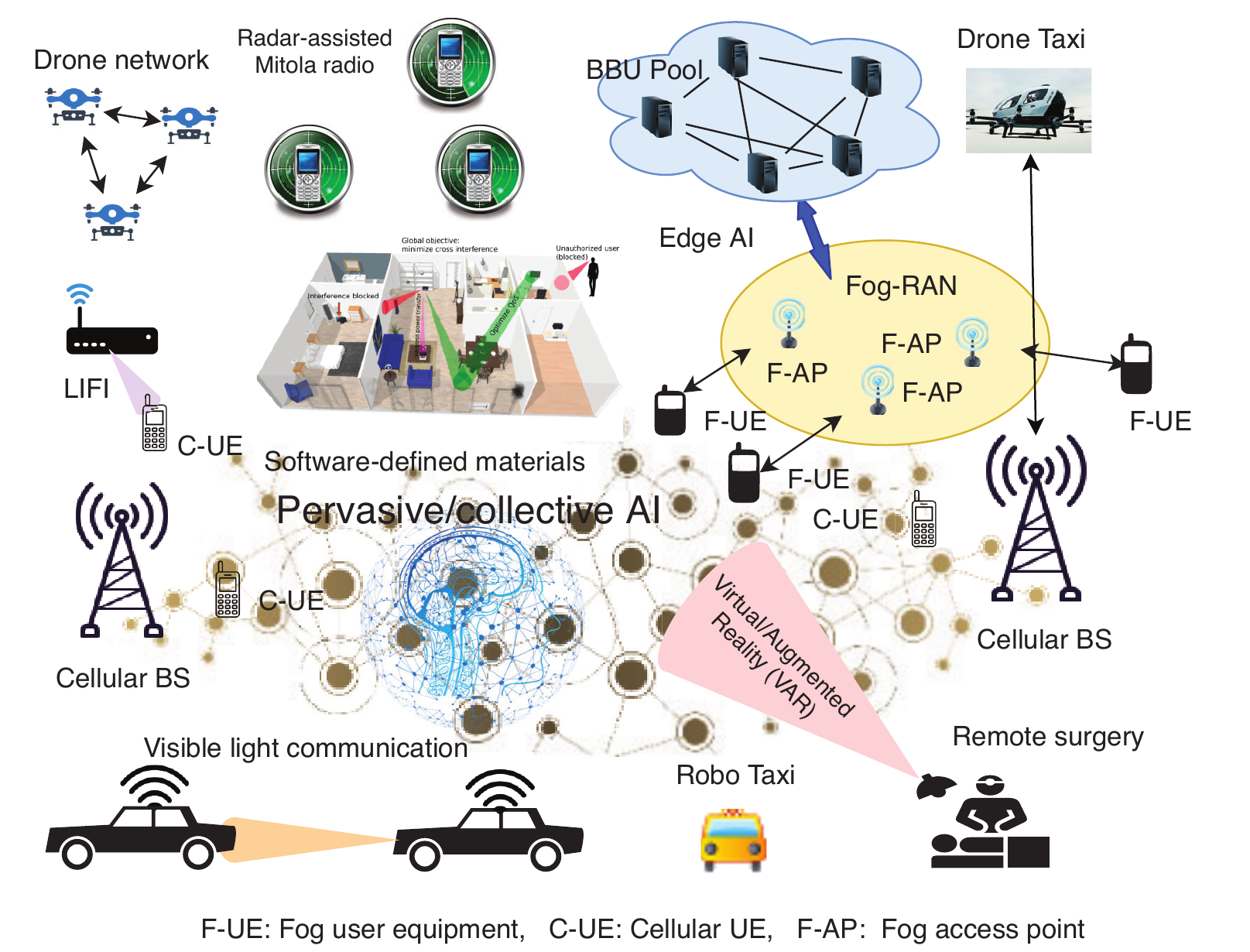}
    \caption{The vision of 6G.}
    \label{fig:vision}
\end{figure}

Fifth-generation (5G) networks are being deployed commercially~\cite{5Gcountries}. However, the rapid growth of data-centric intelligent systems has brought significant challenges to the capabilities of 5G wireless systems. For example, the 5G air interface delay standard of less than $1$ millisecond (ms) is challenging to meet the air interface delay of less than $0.1$ms required by haptic Internet-based telemedicine~\cite{matti2019key}. To overcome the performance limitations of 5G to deal with new challenges, countries are researching the sixth-Generation (6G) mobile communication system.
Upcoming technologies such as artificial intelligence (AI), virtual reality, and the Internet of Everything (IoE) require low latency, ultra-high data rates and reliability. The various applications as shown in Fig.~\ref{fig:vision} cannot be satisfied by existing 5G's ubiquitous mobile ultra-broadband,  ultrahigh data density, and ultrahigh-speed-with-low-latency communications~\cite{chowdhury20196g, saad2019vision, ho2019next, mollah2019emerging}. Performance limitations of 5G and the emerging revolutionary technologies drive the development of 6G networks~\cite{zong20196g}.

Past generations of wireless networks utilize micro-wave communications over the sub-6 GHz band, whose resources are almost used up~\cite{zhu2019millimeter}. Hence, the Terahertz (THz) bands will be the major candidate technology for the 6G wireless communications~\cite{tariq2019speculative, saad2019vision, giordani2019towards, xiao2017millimeter, andrews2016modeling, zhu2019millimeter}. Due to the propagation loss, the THz will be used for high bit-rate short-range communications~\cite{nawaz2019quantum}. Besides, the 90-200GHz spectrum is often not used in the past generations of wireless networks. The sub-THz radio spectrum above 90GHz has not been exploited for radio wireless communications yet; thus, it is envisioned to support the increased wireless network capacity~\cite{corre2019sub}. 6G will undergo the transition from radio to sub-terahertz (sub-THz), visible light communication and terahertz to support explosive 6G applications~\cite{david2019defining}.

\begin{figure}[!h]
    \centering
     \includegraphics[scale=0.3]{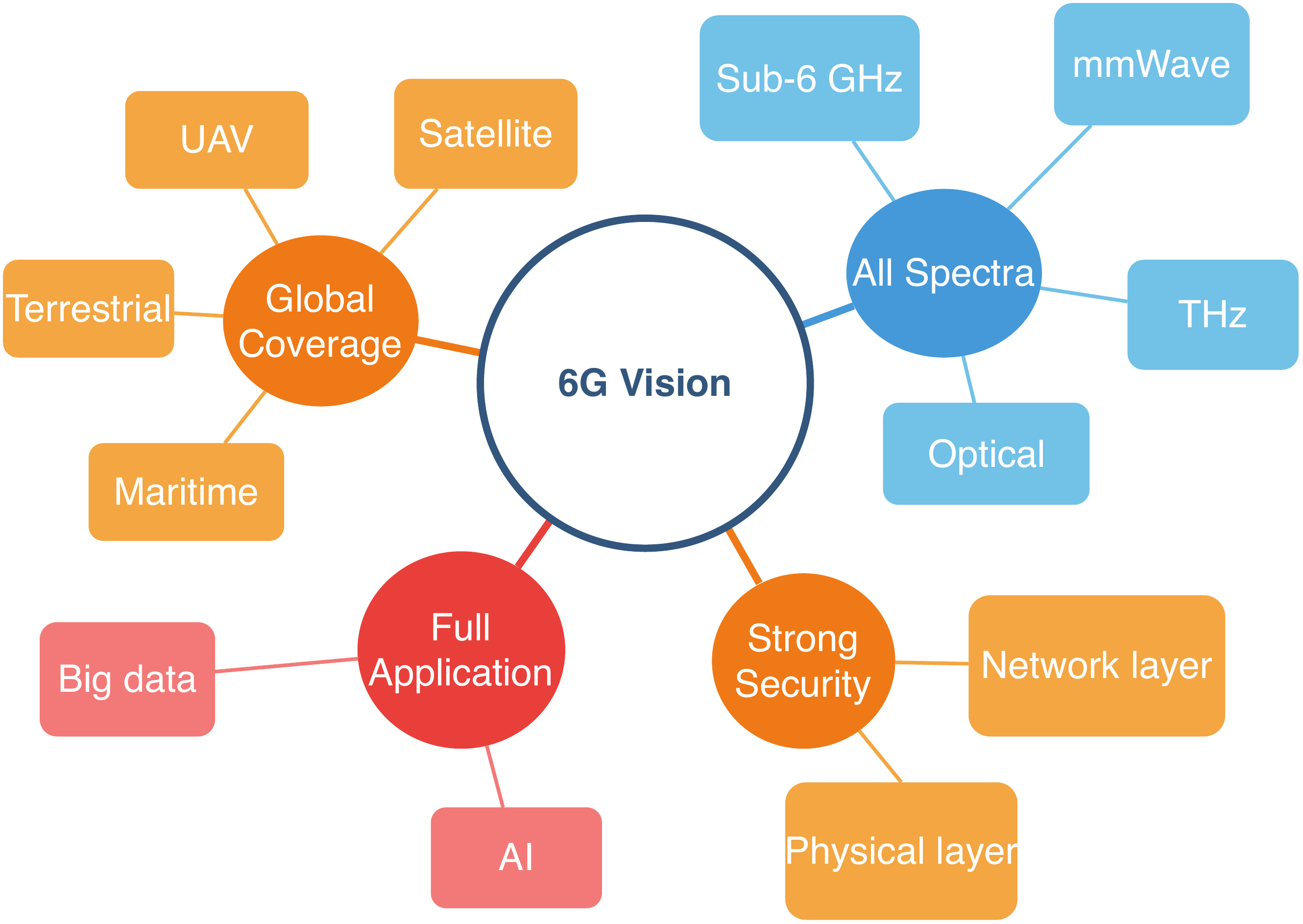}
    \caption{6G wireless communication networks.}\vspace{-10pt}
    \label{fig:6g_vision}
\end{figure}

\begin{table}[!h]
\centering
\scriptsize \setlength{\tabcolsep}{1.1pt} \caption{Requirements and Features of 6G~\cite{ho2019next, piran2019learning,dang2020should,nayak20206g, tariq2019speculative}.}
\label{table}
\begin{tabular}{ll}
\hline
\textbf{Requirements}                       & \textbf{6G}                                                                                                                                                                     \\ \hline
Service types                               & MBRLLC/mURLLC/HCS/MPS                                                                           \\ \hline
Service level                               & Tactile         \\ \hline
Device types                                & Sensors and DLT devices/CRAS/ \\&XR and BCI equipment/Smart implants \\ \hline
Jitters                                     & 1 $\mu$s        \\ \hline
Individual data rate                        & 100 Gbps  \\ \hline
Peak DL data rate                           & $\geq$ 1 Tbps   \\ \hline
Latency                                    & $0.1$ ms                                                                                                                                                                       \\ \hline
Mobility                                    & up to 1000 km/h                                                                                                                                                                 \\ \hline
Reliability                                 & up to 99.99999\%                                                                                                                                                                \\ \hline
Frequency bands                             & \begin{tabular}[c]{@{}l@{}}- sub-THz band\\ - Non-RF, e.g., optical, VLC, laser $\cdots$\end{tabular}                     \\ \hline
Power consumption                           & Ultra low                                                                                                                                                        \\ \hline
Security and privacy                        & Very high                                                                                                                                                                       \\ \hline
Network orientation                         & Service-centric                                                                                                                                                                 \\ \hline
Wireless power transfer \\/Wireless charging & Support (BS to devices power transfer)                                                                                                                                         \\ \hline
Smart city components                       & Integrated                                                                                                                                                                      \\ \hline
Autonomous V2X                              & Fully                                                                                                                                                                           \\ \hline
Localization Precision                      & 1 cm on 3D                                                     \\    \hline
Architecture                &Intelligent Surface \\ \hline
Core network                &Internet of Everything \\ \hline
Satellite integration       & Full  \\ \hline
Operating frequency         &1 THz \\ \hline
Highlight                   &Security, secrecy, privacy \\ \hline
Multiplexing                &Smart OFDMA plus IM  \\ \hline
\end{tabular}
\end{table}
Furthermore, the 6G system is envisioned to support new services such as smart wearable, computing reality devices, autonomous vehicles, implants,  sensing, and 3D mapping~\cite{chowdhury20196g}. 6G's architecture is expected to be a paradigm-shift and carries higher data rates with low latency~\cite{saad2019vision}. Ho~\emph{et~al.}~\cite{ho2019next} and Piran~\emph{et~al.}~\cite{piran2019learning} present their predictions for 6G's requirements and features, which  are summarized in Table~\ref{table}. With these advanced features, 6G wireless communication networks will integrate space-air-ground-sea networks to achieve the global coverage as shown in Fig.~\ref{fig:6g_vision}~\cite{giordani2020satellite}.

\textbf{Contributions of this survey:} In this survey, we focus on the emerging technologies that will be used in the 6G.  The contributions of this survey are summarized below.
\begin{itemize}
    \item Current papers on 6G pay more attention to predicting technologies that may be used in the future, and none of them gives a summary. Our paper surveys existing visions of 6G and summarize them.
    \item We highlight four technologies that are envisioned to play a significant role in 6G, including index modulation, AI, intelligent reflecting surface, and THz communication. 
    \item We briefly explain several potential technologies that have been discussed recently, such as visible light communication, blockchain-based network, satellite communication, holographic radio, and network in box.
\end{itemize}

\textbf{Organization.} The rest of paper is organized as follows. Section~\ref{sec:technologies} introduces emerging technologies that enable the paradigm shift in 6G wireless networks. Section~\ref{sec:conclusion} concludes this paper.

\section{6G Architecture: A Paradigm Shift}~\label{sec:technologies}

In this section, we will explain the most eye-catching ideas pertaining to 6G in detail, including index modulation, artificial intelligence, active/passive intelligent reflecting surfaces and THz communications~\cite{dang2020should, kato2020ten,shafin2020artificial,tan2020thz,elmeadawy2020enabling,gui20206g,chen2020vision,xiaohutowards}. Also, we will present some promising technologies such as blockchain, satellite communication, full-duplex, and holographic radio in short.

\subsection{Index Modulation}

Index Modulation (IM) has high spectral efficiency and high power efficiency due to its idea of sending extra information through the indexed resource entities such as the time slots, the transmit/receive antennas, the subcarriers and the channel states. Its low deployment cost and high throughput attract much attention in the upcoming 6G communications.

\begin{figure*}[htbp]
    \centering
    \includegraphics[scale=0.8]{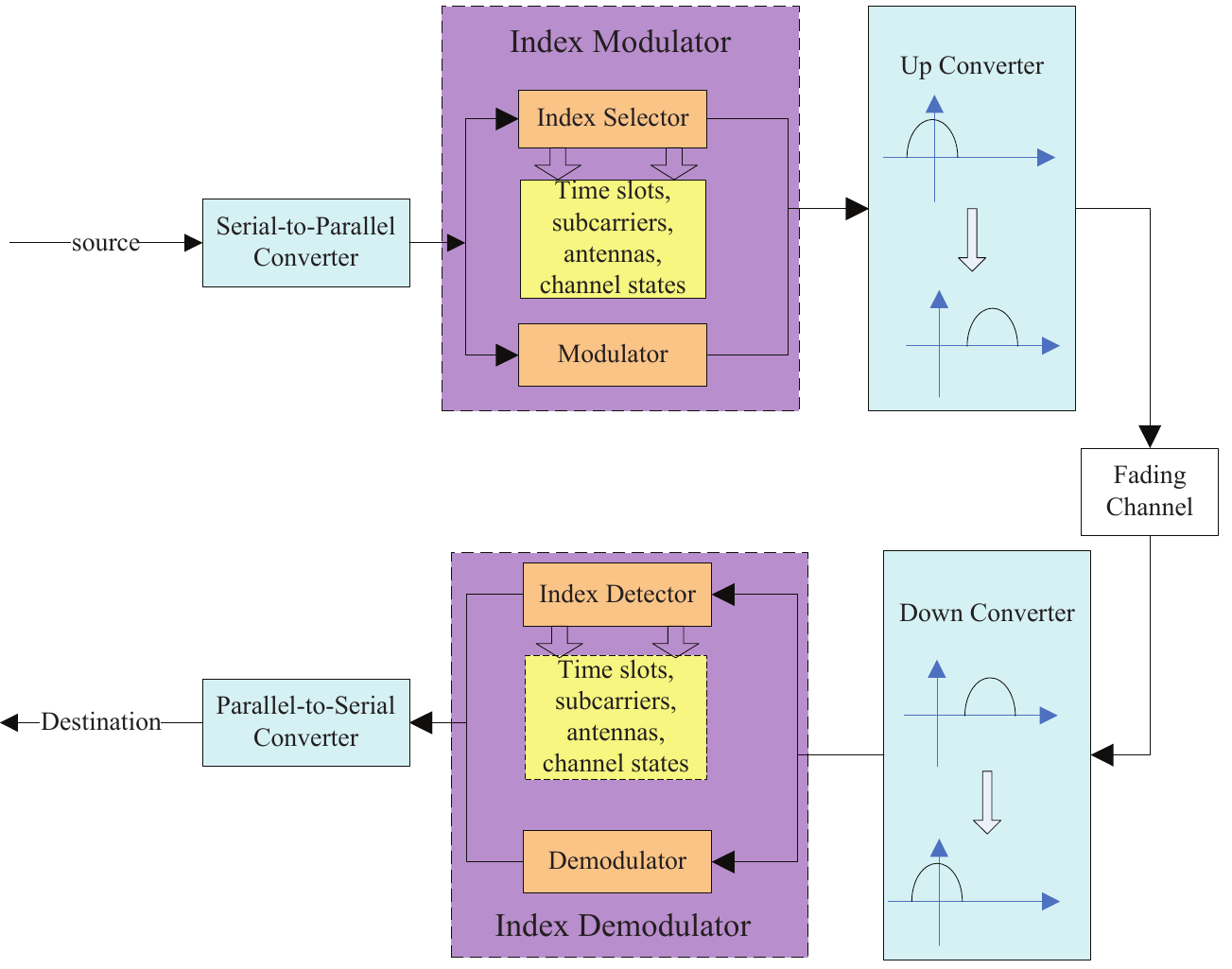}
    \caption{The structure of IM-aided systems.}\vspace{-10pt}
    \label{fig:IM}
\end{figure*}

Based on the entities of the indexed resources, IM can be classified into time-domain IM, spatial-domain IM, frequency-domain IM,  and channel-domain IM. Each kind of IM technique divides the transmitted bits into two parts, as illustrated in Fig.~\ref{fig:IM}: one part for classical modulation, for example, phase shift keying and quadrature amplitude modulation,  etc.; the other part for the activation of the indexed resources used for transmitting the additional bits. Because the additional information requires neither spectrum resources nor power resources, IM can improve the throughput while consuming low power compared with its non-IM-aided counterpart.

\noindent\textbf{TD-IM Technique.} Time Division Duplex (TDD) is one of the generally used kind of wireless techniques, and will be inherited in 6G communications without doubt. In TDD transmission, a data frame consists of several time slots, and each time slot can be used to transmit the source information. However, for TD-IM technique, only a fraction of time slots activated by the transmitting bits are used for signal transmission. To further improve the demodulation performance, space-time shift keying (STSK) technique is proposed by Sugiura \emph{et al.} in~\cite{SugiuraSTSK} to combine TD-IM and space-time block code (STBC) together where a dispersion matrix can be activated for index selection. Subsequently, the same authors extend STSK to a general format in~\cite{SugiuraGSTSK}, where indices of multiple activated dispersion matrices are chosen based on the transmission information bits, further improving the capacity.

\noindent\textbf{FD-IM Technique.} The orthogonal frequency division multiplexing (OFDM) technique is widely used due to its high spectral efficiency in 4G and 5G communications and will still be used in the 6G network. The bandwidth is divided into several subcarriers orthogonally, and each subcarrier transmits its own data bits individually. For the FD-IM technique, additional bits are used to choose the indices of the activated subcarriers. Therefore, FD-IM is also called subcarrier-indexed OFDM (SIM-OFDM). Different from TD-IM, the FD-IM technique has much more resource entities. Thus, FD-IM often divides the subcarriers into several blocks, and activates one subcarrier (SIM-OFDM)~\cite{AlhigaSIM_OFDM} or more than one subcarriers (generalized SIM-OFDM, GSIM-OFDM)~\cite{RanGSIM_OFDM} in each subblock, reducing the number of index patterns at the transmitter as well as the demodulation complexity at the receiver. When combing with MIMO systems, SIM-OFDM can also be extended to MIMO-OFDM-IM to achieve considerable performance gain \cite{BasarMIMO_OFDM_IM, BasarMIMO_OFDM_IM1}.

\noindent\textbf{SD-IM Technique.} SD-IM is also called spatial modulation (SM), where the spare information bits are used to activate the transmit antennas~\cite{YangSM}. Compared with MIMO systems, SM needs no inter-antenna synchronization and is free of inter-antenna interference, which leads to low receiver complexity. The SM can also be extended to general SM (GSM) to improve the data rate when multiple antennas are activated to transmit the modulated signals. On the other hand, precoding SM (PSM) or general precoding SM (GPSM) can be applied by activating the receive antennas to exploit the advantage of beamforming~\cite{ZhangPSM, LiuPSM}. In PSM/GPSM, a precoding scheme is implemented at the transmitter to identify the desired antennas/antennas, thus providing transmit diversity and improving the detect performance at the receiver.

\noindent\textbf{CD-IM Technique.} Unlike the aforementioned IM schemes, CD-IM can change the property of radio frequency (RF) environment by employing RF mirrors or electronic switches~\cite{SeifiMBM, BouidaMBM, NareshMBM, BasarMBM}. Therefore, CD-IM has also been named media based modulation (MBM). MBM uses several RF mirrors/electronic switches around the transmit antennas. It allows the signal to transmit to the receiver through distinct channel paths according to the on/off status of the RF mirrors/electronic switches. Compared with the MIMO system, where the channel matrix is generally nonorthogonal, MBM can further randomize the channel by perturbing the wireless environment, enhancing the achievable rate. MBM can be combined with MIMO systems (MIMO-MBM)~\cite{SeifiMBM, BouidaMBM}, Alamouti STBC systems (STCM)~\cite{BasarMBM}, etc., to improve the capacity further or detect performance.

In addition, there are some other complex IM schemes. In~\cite{ShamasundarMBM}, Schamasundar \emph{et al.} propose a scheme by combining the TD-IM and MBM, namely, time-indexed MBM. Such hybrid IM technique can increase the data rate considerably at the cost of detect complexity. In~\cite{ErtugrulMBM}, Ertugrul \emph{et al.} give a novel sparse code multiple access (SM-SCMA) scheme operating in uplink transmission, which is used for organizing the accessing of multiple users. SM-SCMA is an example of non-orthogonal multiple access aided SM (NOMA-SM), which is remarkable in reducing the inter-user interference in 6G multi-user communications~\cite{WangNOMA_SM}. IM-aided systems can also be designed with bit-interleaved coded modulation (BICM) where soft information between the channel decoder and index pattern can be exchanged iteratively to obtain near-capacity performance~\cite{WangBICM}. In~\cite{XiaoCS}, a reduced complexity detector is presented by introducing the compressing theory into the IM-aided system, where the sparsity of the activated resource entities is exploited to detect the information.

\subsection{Artificial Intelligence}
\begin{figure*}[htb]
    \centering
    \includegraphics[scale=0.55]{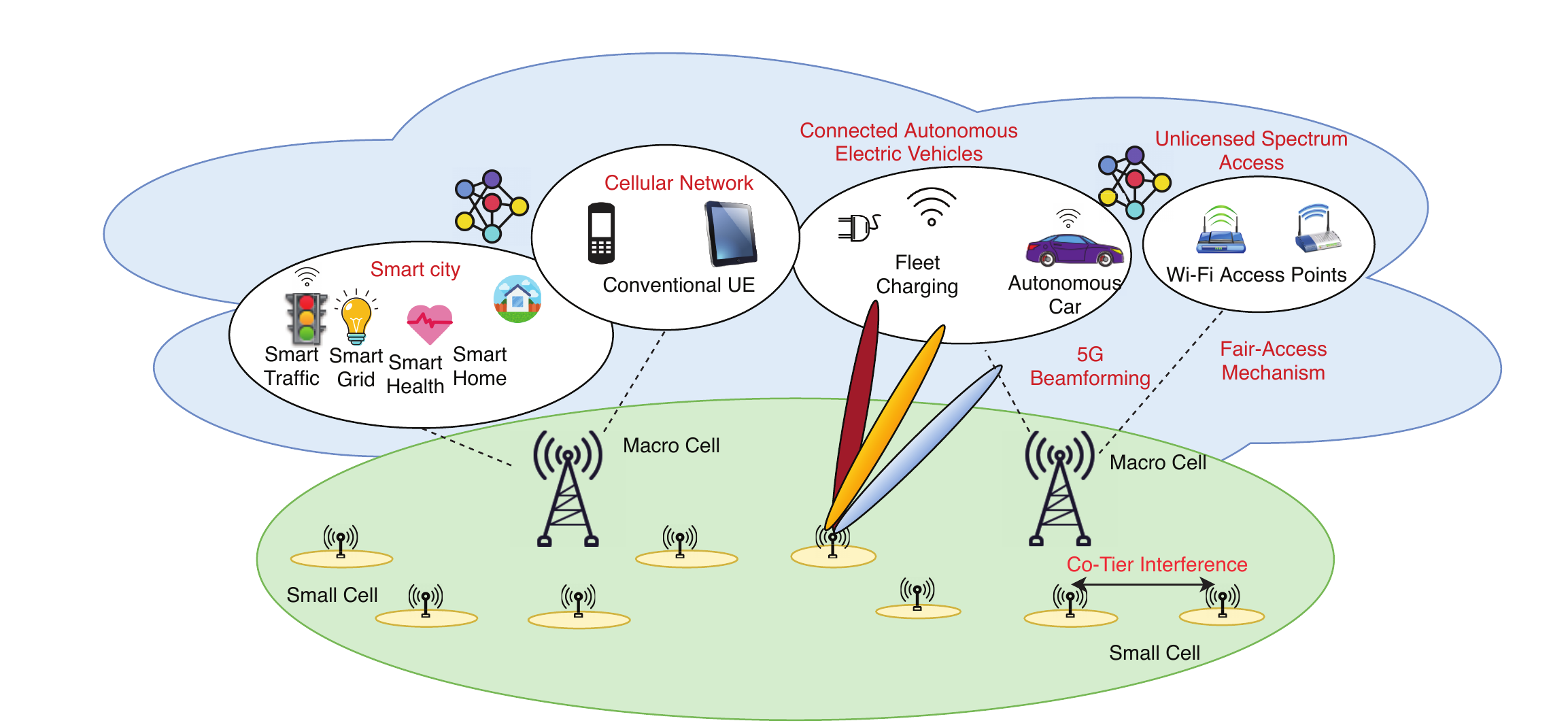}
    \caption{An AI-Enabled 6G wireless network and related applications.} \vspace{-10pt}
    \label{fig:AI}
\end{figure*}

Artificial Intelligence (AI) provides intelligence for wireless networks by simulating some human thought processes and intelligent behaviors.  By leveraging AI, 6G will enable more applications illustrated in Fig.~\ref{fig:AI} to be intelligent such as smart city, cellular network, connected autonomous electric vehicles, and unlicensed spectrum access~\cite{david2019defining, zong20196g, letaief2019roadmap, elsayed2019ai, zappone2019model, gacanin2019autonomous,  strinati20196g, stoica20196g, zhang20196g, zhangLin20196g, piran2019learning}. In particular, AI, residing in new local ``clouds" and ``fog" environments, will help to create many novel applications using sensors that will embedded into every corner of our life~\cite{giordani2019towards}. In the following, we present some essential AI techniques and corresponding applications for 6G.

Deep learning, considered as the vital ingredient of AI technologies, has been widely used in the wireless networks~\cite{mao2018deep}. It will play an essential role in various areas, including, semantic communications, holistic management of communication, control resources areas,  caching, and computation, etc., which push the  paradigm-shift of 6G.

\subsubsection{Artificial Intelligence Algorithms.}

In this section, we summarize some potential AI techniques: supervised and unsupervised learning, model-driven deep learning, deep  reinforcement  learning, federated learning, and explainable  artificial  intelligence as follows.

\noindent\textbf{Supervised Learning.} The supervised Learning trains the machine model using labelled training data~\cite{piran2019learning}. There are some well developed algorithms that can be used in the 6G network, such as support vector machines, linear regression, logistic regression, linear discriminant analysis, naive Bayes, k-nearest neighbors and decision tree, etc. Supervised learning techniques can be used in both physical layer and network layer. In physical layer, we can utilize supervised learning for channel states estimation, channel decoding, etc.  Supervised learning techniques can be deployed for caching, traffic classification, and delay mitigation and so on in the network layer.

\noindent \textbf{Unsupervised Learning.}
Unsupervised learning is leveraged to find undefined patterns in the dataset without using labels. Commonly used unsupervised learning techniques include clustering, anomaly detection, autoencoders, deep belief nets, generative adversarial networks, and expectation–maximization algorithm. At the physical layer, unsupervised learning techniques are applicable to optimal modulation, channel-aware feature-extraction,  etc. In addition, unsupervised learning technologies can be used for routing, traffic control and parameter prediction, etc, in network layer.

\noindent\textbf{Model-Driven Deep Learning.} The model-driven approach is to train an artificial neural network (ANN) with prior information based on professional knowledge~\cite{zappone2019model, gacanin2019autonomous, he2019model}. The model-driven approach is more suitable for most communication devices than the pure data-driven deep learning approach, because it does not require  tremendous computing resources and considerable time to train what the data-driven method needs~\cite{he2019model}. The approach to apply model-driven deep learning proposed by Zappone~\emph{et~al.}~\cite{he2019model} includes two steps:  first, we can use theoretical models derived from wireless communication problems as prior expert information. Secondly, we can subsequently tune ANN with small sets of live data even though initial theoretical models are inaccurate.

\noindent\textbf{Deep Reinforcement Learning.}
Deep reinforcement learning (DRL) leverages Markov decision models to select the next ``action" based on the state transition models~\cite{mao2018deep}. DRL technique is considered as one of the promising solutions to maximize some notion of cumulative reward by sequential decision-making~\cite{piran2019learning}. It is an approach to solve resource allocation problems in 6G~\cite{elsayed2019ai, zhangLin20196g}. As 6G wireless networks serve a wider variety of users in the future, the radio-resource will become extremely scarce. Hence, efficient radio-resource allocation is urgent and challenging~\cite{elsayed2019ai}.

\noindent\textbf{Federated Learning.}
Federated Learning (FL) aims to  train a machine  learning model with training data remaining distributed at clients in order to protect data owners' privacy~\cite{konevcny2016federated}. As 6G heads towards a distributed architecture, FL technologies can contribute to enabling the shift of AI moving from a centralized cloud-based model to the decentralized devices based~\cite{letaief2019roadmap, shafin2019artificial,tariq2019speculative}. In addition, since the edge computing and edge devices are gaining popularity, AI computing tasks can be distributed from a central node to multiple decentralized edge nodes. Thus, FL is one of the essential machine learning methods to enable the deployment of accurately generalized models across multiple devices~\cite{cousik2019cogrf}.

\noindent\textbf{Explainable Artificial Intelligence.}
Since there will emerge a large scale of applications such as remote surgery and self-driving in the 6G era, it is necessary to make  artificial intelligence explainable for building trust between humans and machines.  Currently, most AI approaches in PHY and MAC layers of 5G wireless networks are inexplicable~\cite{guo2019explainable}.  AI applications such as self-driving and remote surgery are considered to be widely used in 6G, which requires explainability to enable trust. AI decisions should be explainable and understood by human experts to be considered as trustworthy. Existing methods, including visualization with case studies, hypothesis testing, and didactic statements, can improve deep learning explainability

\subsubsection{Artificial Intelligence Applications in 6G.}

In this section, we present some potential use cases of AI in 6G such as AI in network management and AI in autonomy.

\noindent\textbf{AI in network management.} As 6G network becomes complex, it may utilize deep learning instead of human operators to improve the flexibility and efficiency in the network management~\cite{piran2019learning}. AI technologies are applicable to both the physical layer and network layers. In physical layer, AI techniques have involved in design and resource allocation in wireless communications~\cite{ho2019next}.  For example, unsupervised learning are applicable to interference cancellation, optimal modulation, channel-aware feature-extraction, and channel estimation, etc.~\cite{piran2019learning}. Deep reinforcement learning is possible to be employed for link preservation, scheduling,  transmission optimization, on-demand beamforming, and energy harvesting, etc.~\cite{piran2019learning, shafin2019artificial}. In addition, AI technologies can be used to the network layer as well. Supervised learning techniques can tackle problems such as resource allocation, fault prediction, etc.~\cite{piran2019learning}. Besides, unsupervised learning algorithms can help in routing, traffic control, parameter prediction, resource allocations, etc.~\cite{piran2019learning}. Reinforcement learning can be important for traffic prediction, packet scheduling, multi-objective routing, security, and classification, etc.~\cite{piran2019learning, shafin2019artificial}.

\noindent\textbf{AI in Autonomy.}
AI technologies are potential to enable 6G wireless systems to be autonomous~\cite{loven2019edgeai,gacanin2019autonomous,zhang20196g}. Agents with intelligence can detect and resolve network issues actively and autonomously. AI-based network management contributes to  monitoring network status in real-time and keep network health. Also, AI techniques can provide intelligence at the edge devices and edge computing, which enables edge devices and edge computing to learn to solve security problems autonomously~\cite{porambagesec, mollah2017secure, loven2019edgeai}. In addition, autonomous applications such as autonomous aerial vehicles and autonomous robots are envisioned to be available in 6G~\cite{ho2019next}.



\begin{figure*}[htbp]
    \centering
    \includegraphics[scale=0.12]{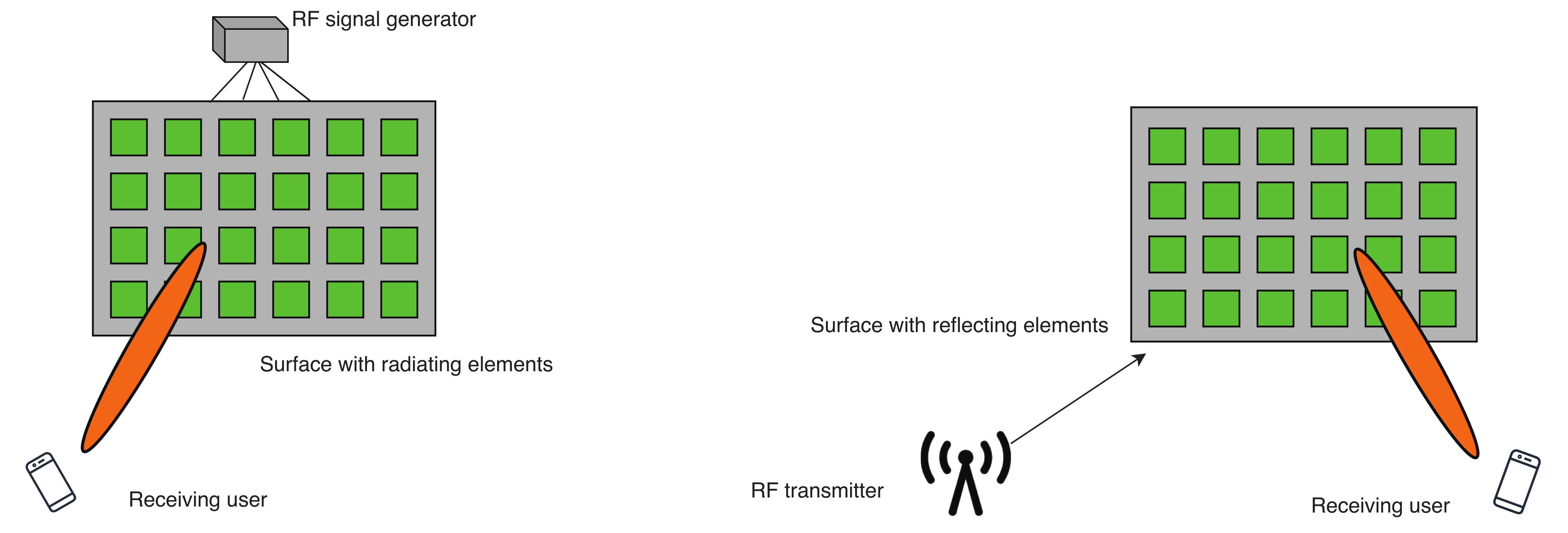}\vspace{-10pt}
    \caption{Left: Large Intelligent Surfaces (an RF signal generator locates at the backside). Right: Intelligent Reflecting Surfaces (an RF signal generator locates at another location).}\vspace{-20pt}
    \label{fig:intelligent_surfaces}
\end{figure*}

\subsection{Intelligent Surfaces}

Currently, two types of intelligent surfaces shown in Fig.~\ref{fig:intelligent_surfaces} attract researchers' attention - Large Intelligent Surfaces (LISs) and Intelligent Reflecting Surfaces (IRSs). LISs are useful for constructing  an intelligent and active environment with integrated electronics and wireless communications~\cite{faisal2019ultra, hu2018beyond}. Renzo~\emph{et~al.}~\cite{di2019smart} believe that IRSs will be utilized in 6G, because they predict that future's wireless networks will serve as an intelligent platform connecting the physical world and the digital world seamlessly. They foresee that wireless networks will be smart radio environments which are potential to realize uninterrupted wireless connectivity and use existing radio waves to transmit data without generating new signals.

\subsubsection{Large Intelligent Surfaces (LISs).}
The concept of deploying antenna arrays as LISs in massive MIMO systems was originally proposed by Hu~\emph{et~al.}~\cite{hu2017potential}. LISs are electromagnetically active in the physical environment, where each part of an LIS can send and receive electromagnetic fields. Buildings, streets, and walls are expected to be electronically active after decorating with LISs~\cite{faisal2019ultra}. LISs have the following main favorable features~\cite{faisal2019ultra}:
(i) Generate perfect LoS indoor and outdoor propagation environments.
(ii) They put little restriction on the  spread of antenna elements. Therefore, antenna correlations and  effects of mutual coupling can be avoided more easily, such that sub-arrays are large and the channel is well-conditioned for propagation. Thus, LISs can be realized via THz Ultra-Massive MIMO (UM-MIMO). LISs are very useful for applications with low-latency, because channel estimation techniques and feedback mechanisms that LISs support are simple.

\subsubsection{Intelligent Reflecting Surfaces (IRSs).}

The IRS is considered as a promising candidate in improving the quality of the signal at the receiver by modifying the phase of incident waves~\cite{hu2018beyond, nadeem2019large, jung2019performance, de2019non, hu2017potential, hu2018user, hu2017cramer,gong2019towards}. IRSs are made of electromagnetic (EM) material that are electronically controlled with integrated low-cost passive reflecting elements, so  that they contribute to forming the smart radio environment~\cite{basar2019wireless}. IRS can change the wireless signal propagation environment by adjusting the phase shift of the reflecting elements. Besides, IRSs help to enhance the communication between a sender and a receiver by reflecting the incident wave~\cite{basar2019wireless, ozdogan2019intelligent, liang2019large}. By adjusting the reflection coefficients, IRSs enable the reflected signals being coherently added to the receiver without adding additional noise~\cite{liang2019large}. Besides, IRSs can increase signal power and modify signal phase~\cite{faisal2019ultra}. In particular, by utilizing local tuning, graphene-based plasmonic reconfigurable metasurfaces can obtain  some benefits, including beam focusing, beam steering, and  control on wave vorticity~\cite{liaskos2018new}. Unlike LISs, IRSs use passive array architecture for reflecting purpose~\cite{qingqing2019towards}. Distinguishable features of IRSs summarized by Basar~\emph{et~al.}~\cite{basar2019wireless} and Wu~\emph{et~al.}~\cite{qingqing2019towards} include:
\begin{itemize}
    \item They comprise low-cost passive elements which are controlled by the software programming.
    \item They do  not  require specific  energy  source to support during  transmission.
    \item They do not need any backhaul connections to exchange traffics.
    \item The IRS is  a configurable surface, so that points on its surface can shape the wave impinging upon it.
    \item They are fabricated with  low profile, lightweight, and conformal geometry such that they can be easily deployed.
    \item They work in the full-duplex mode.
    \item No self-interference.
    \item The noise level does not increase.
\end{itemize}

Different from existing technologies such as  backscatter communication, active relay, and active surface based massive MIMO, the IRS-aided network includes both active components (BS, AP, user terminal) and passive component (IRS). We highlight some differences between IRS and well-known technologies as follows.

\noindent \textbf{Massive MIMO.} IRSs and massive MIMO consist of different array architectures (passive versus active) and operating mechanisms (reflecting versus transmitting)~\cite{qingqing2019towards}. Benefit from the passive elements, IRSs achieve much more gains compared to massive MIMO while consuming low energy~\cite{yuan2019potential}.

\noindent\textbf{Amplify-and-Forward (AF) Relay.} Relay uses active transmit elements to assist the source-destination communication, but the IRS serves as a passive surface, which reflects the received signal~\cite{qingqing2019towards,yuan2019potential}. Relays help to reduce the rate of the available link  if they are in half-duplex mode.  When  they operate in full-duplex mode, they suffer the severe self-interference. 
Active relay usually works in half-duplex mode for reduced self-interference. While IRS can work in full-duplex mode, improving the spectrum efficiency compared to the former. Active relay usually works in the half-duplex mode, which wastes spectrum compare to IRS, which works under full-duplex mode. If AF implements full-duplex mode, it will require costly self-interference cancellation techniques to support. But IRS overcomes above outstanding shortcomings of AF relays.

\noindent \textbf{Backscatter.} Backscatter requires the reader to realize self-interference cancellation at the receiver to decode the radio frequency identification (RFID) tag's message. RFID communicates with the reader by modulating its reflected signal sent from the reader~\cite{qingqing2019towards}. However, IRS only reflects received signals without modifying information; thus, the receiver can add both the direct-path and reflect-path signals to improve the decoding's signal strength.

\subsection{Terahertz Communications}
Currently, wireless communication systems are unable to catch up with the ever-increasing number of applications in 6G. Terahertz (THz) frequency band, which ranges from 0.1 to 10 THz, is the unexplored span of radio spectrum~\cite{yuan2019potential, sarieddeen2019next}. THz communications provide new communication paradigms with ultra-high bandwidth and ultra-low latency~\cite{sarieddeen2019next}. It is envisioned the data rate should be as high as Tbps to satisfy 6G applications' requirements of high throughput and low latency~\cite{yuan2019potential}. A novel approach to generate the THz frequency is discovered by Chevalier~\emph{et al.}~\cite{chevalier2019widely}. They build a compact device that can use the nitrous oxide or laughing gas to produce a THz laser. The frequency of the laser can be tuned over a wide range at room temperature.  Traditionally, the THz frequency band limits the widespread use of THz. THz transceiver design is regarded as the most critical factor in facilitating THz communications~\cite{yuan2019potential}.

\subsubsection{Terahertz Source Technique.} 
Recent technology advancements in THz transceivers, such as photonics-based devices and electronics-based devices, overcome the THz gap, and enable some potential use cases in 6G~\cite{sarieddeen2019next}. The electronic technologies such as silicon-germanium BiCMOS, III-V semiconductor,  and standard silicon CMOS related technologies (III and V represents the old numbering of the periodic system groups), have been vastly advanced, such that mixers and amplifiers can operate at around 1THz frequency~\cite{yuan2019potential,han2019terahertz}. The photonic technologies are possiblely be used in the practical THz communication systems~\cite{yuan2019potential,han2019terahertz}. In addition, the combination of electronic-based transmitter and photonics-based receiver is feasible. Recent nanomaterials may help to develop novel devices that can used for THz communications~\cite{han2019terahertz}.

\subsubsection{Applications of Terahertz.}
Due to the high transmission RF, signals transmitted through THz frequency band suffer from a high pass loss. According to the Friis' law, the pass loss in free space increases quadratically with the operating frequency~\cite{HanTHz}. This feature limits the use of THz to short-distance transmission such as indoor communications~\cite{KhalidTHz}. Meanwhile, THz band can satisfy the requirement of ultra-high data rate; therefore, ultra-broadband applications, for example, virtual reality (VR) and wireless personal area networks can also exploit THz band to transmit signals~\cite{HanTHz}. THz technique can also be used in secure wireless communications. Since THz signals possess a narrow beam, it's very hard  to wiretap the information for the eavesdropper when locating outside the transmitter beam ~\cite{KhalidTHz}.

\subsection{Other Potential Technologies}

In this section, we introduce some potential technologies that will be used in 6G, for example, visible light communications, blockchain-enabled network, satellite communication,  full-duplex technology, holographic radio, and network in box.

\subsubsection{Visible Light Communications.} 

Visible light communication (VLC) is considered as one of the techniques that will be used in  6G communications. VLC contains transmitters and receivers. For short-range communication, either data-modulated white laser diodes or light-emitting diodes are used  as transmitters, while photodetectors are utilized as receivers. Besides, VLC is considered as a complementary technology of RF communications since it can utilize an unlicensed spectrum for communication~\cite{giordani2019towards}.


The laser diode (LD)-phosphor conversion lighting technology can provide better performance in efficiency and brightness,  and larger illumination range compared with traditional lighting techniques~\cite{yuan2019potential}. Thus, it is considered as the most promising technology for 6G. The speed LD-based VLC system is possible to reach 100Gbps, which meets the requirements of  ultra-high data density (uHDD) services in 6G. Besides, the upcoming new light sources based on microLED will overcome the limitation of low speed in short range communication~\cite{strinati20196g}. As massive parallelization of microLED arrays, spatial multiplexing techniques, CMOS driver arrays, and THz communications develop, VLC's data rate is expected to reach Tbps  in the short range indoor scenario by the year of 2027~\cite{strinati20196g, sarieddeen2019next}.

VLC can be used in indoor scenarios because it has limited coverage range, and it needs an illumination source and suffers from interference from other sources of light (e.g., the sun)~\cite{giordani2019towards}. Also, the space-air-terrestrial-sea integrated network can use VLC to provide better coverage~\cite{nawaz2019quantum}. In addition, traditional electromagnetic-wave signals cannot achieve high data transmission speed using laser beams in the free space and underwater, but VLC has ultra-high bandwidth and high data transmission speed~\cite{chowdhury2019role}. Therefore, VLC are useful in cases where traditional RF communication is less active, for example, in-cabin internet service~\cite{tariq2019speculative}. Furthermore, VLC is envisioned to be widely used in vehicle-to-vehicle communications, which depend on the head and tail lights of cars  for communications~\cite{tariq2019speculative, chowdhury2019role, yuan2019potential}. Besides, VLC serves as a potential solution to build gigabit wireless networks underwater.

\subsubsection{Blockchain-based network.} 
\begin{figure}[!h]
    \centering
    \includegraphics[scale=0.6]{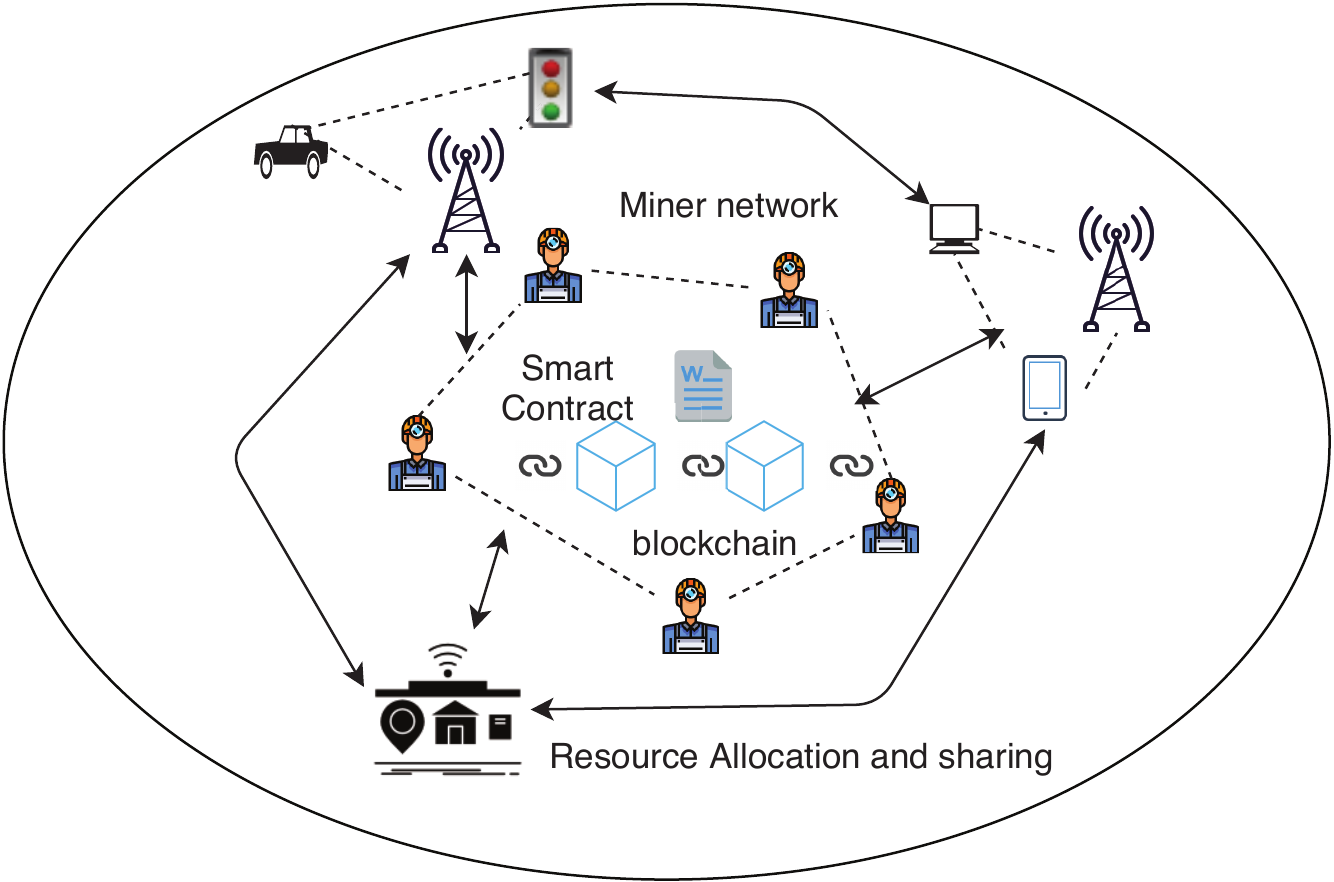}
    \caption{Blockchain-based network.}\vspace{-10pt}
    \label{fig:blockchain}
\end{figure}

Blockchain is a chain of blocks which constitute a distributed database. It is designed for cryptocurrencies (e.g. bitcoin) initially. However, nowadays, blockchain can do more than just in cryptocurrencies but run Turing-complete programs such as smart contracts in a distributed way (e.g. Ethereum)~\cite{wood2014ethereum}. Blockchain provides a distributed and secure database for storing records of transactions, and each node includes the previous block's cryptographic hash, transaction data, and a time stamp~\cite{mollah2020blockchain, zhang20196g}.  Besides, blockchain-like mechanisms are expected to provide distributed authentication, control by leverging digital actions provided by the smart contracts~\cite{strinati20196g}. Combining with federated learning, blockchain-based AI architectures are shifting AI processing to the edge~\cite{porambagesec}.  Recently, a blockchain radio access network (B-RAN) has been proposed with prototype~\cite{ling2019blockchain, le2019prototype}. Thus, blockchain can help to form a secure and decentralized environment in 6G. Blockchain can provide a secure architecture for 6G wireless networks as shown in Fig.~\ref{fig:blockchain}~\cite{dai2019blockchain}. 

\subsubsection{Satellite Communication.} 
\begin{figure}[!h]
    \centering
    \includegraphics[scale=0.55]{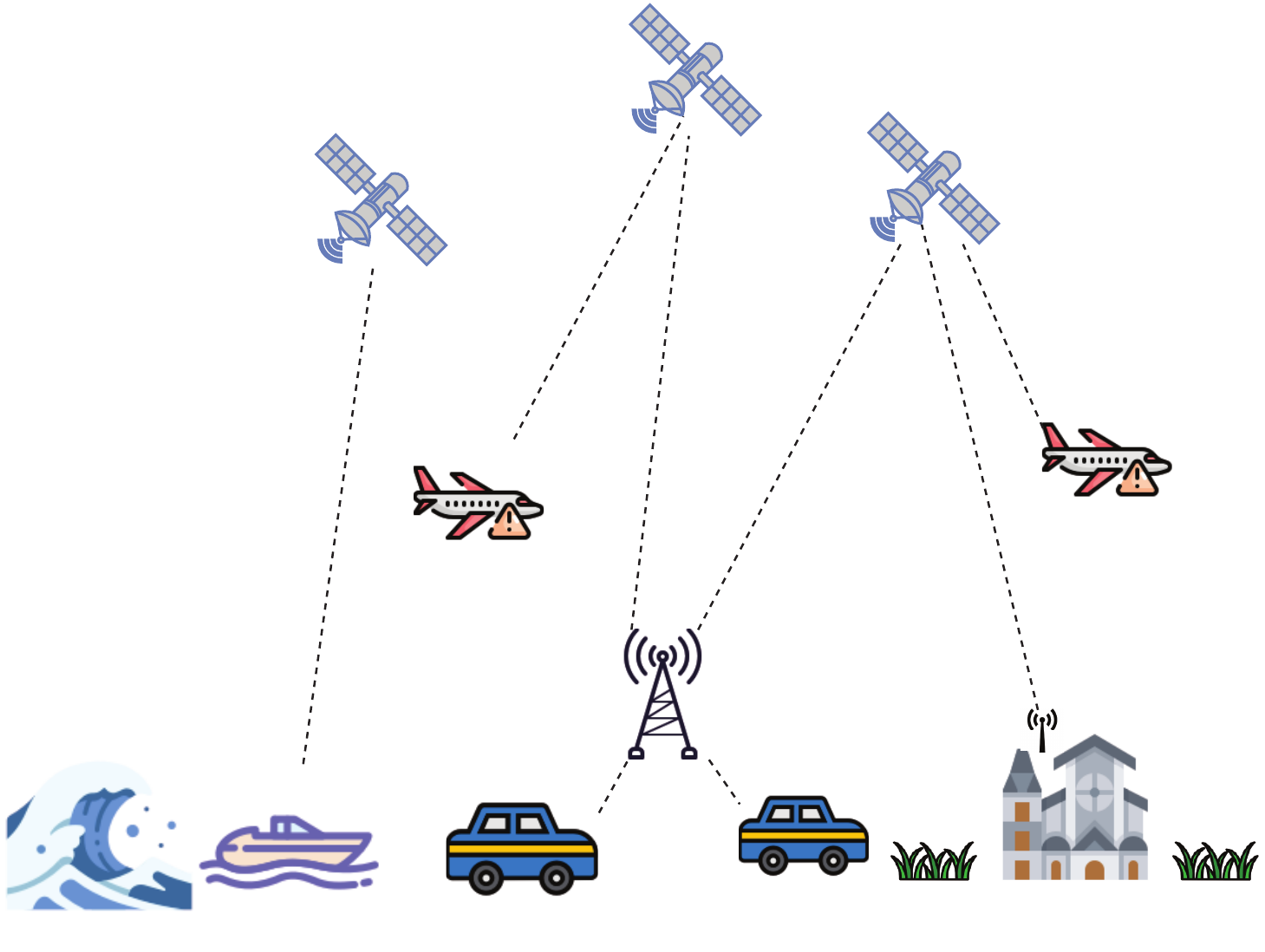}\vspace{-10pt}
    \caption{Satellite communication.}
    \label{fig:satellite}
\end{figure}

Satellite communication means that earth stations communicate with each other via satellites. Satellite communication is a promising solution to the global coverage (i.e. space-air-ground-sea integrated network) in 6G era as shown in Fig.~\ref{fig:satellite}. By integrating with satellite communication, 6G can provide localization services, broadcast, Internet connectivity, and weather information to cellular users~\cite{piran2019learning}. 

\vspace{-10pt}

\subsubsection{Full-duplex Technology.} 

Full-duplex and in-band full-duplex (IBFD) technologies improve the communication efficiency by allowing devices to transmit and receive a signal in the same frequency band~\cite{rajatheva2020white}. Full-duplex technologies are able to make the current efficiency of sharing spectrum double and increase the networks and communication systems' throughput. In the 4G/5G wireless systems, transmission and reception have to be done at the different frequency bands because half-duplex such as frequency division duplex (FDD) or time division duplex (TDD) do not support performing the transmission and reception at the same time slot. In addition, compared with half-duplex, full-duplex technology leverages self-interference cancellation technology to increase the utility of spectral resources, reduce the transmission delay, and improve the throughput between transceiver and receiver links. The hard part is to eliminate the self-interference, and now transmit signal generates over 100dB higher noise than the receiver noise floor. Thus, the new scheduling algorithms should be designed. Currently, three types of self-interference cancellation techniques are proposed, including digital cancellation, analog cancellation, and passive suppression.

\subsubsection{Holographic radio.} Unwanted signals are treated as harmful interference in traditional wireless networks, but they are considered as useful resources to develop holographic communication systems~\cite{zong20196g}. Computational holographic radio is one of the most promising interference-exploiting technologies~\cite{yuan2019potential}.

\subsubsection{Network in box (NIB).} More and more kinds of technologies will be embedded in the 6G wireless network, such as autonomous vehicles, factory automation, etc. To satisfy the real-time and reliable features of the network, the NIB technique is paid much attention to industrial automation because NIB offers a device that can provide seamless connectivity between different services. On the other hand, NIB covers the wireless environments, including terrestrial, air, and marine well agreed to the vision of Internet of Everything in 6G network.

\section{Conclusion}~\label{sec:conclusion}
In this paper, we highlight some promising technologies in 6G networks. We present a detailed explanation of artificial intelligence, intelligent reflecting surfaces, and THz communications. Furthermore, we briefly introduce some promising technologies, including blockchain, satellite communication, full-duplex, holographic radio, and THz communication. We envision that industry and academia will pay more attention to these technologies in 6G. The aforementioned technologies will contribute to 6G in the future. \\

\noindent\textbf{Acknowledgments.} The research is supported by 1) Nanyang Technological University (NTU) Startup Grant, 2) Alibaba-NTU Singapore Joint Research Institute (JRI), 3) Singapore Ministry of Education Academic Research Fund Tier 1 RG128/18,  Tier 1 RG115/19, Tier 1 RG24/20, Tier 1 RT07/19, Tier 1 RT01/19, and Tier 2 MOE2019-T2-1-176, 4) NTU-WASP Joint Project, 5)  Energy Research Institute @NTU (ERIAN), 6) Singapore NRF National Satellite of Excellence, Design Science and Technology for Secure Critical Infrastructure NSoE DeST-SCI2019-0012, 7) AI Singapore (AISG) 100 Experiments (100E) programme, and 8) NTU Project for Large Vertical Take-Off \& Landing (VTOL) Research Platform.

\bibliographystyle{IEEEtran}
\bibliography{related}
\end{document}